\documentclass[a4paper,fleqn,usenatbib]{mnras}

\usepackage{newtxtext,newtxmath}

\usepackage[T1]{fontenc}
\usepackage{ae,aecompl}

\usepackage{hyperref}
\usepackage{chngpage}
\usepackage{soul}
\usepackage{ulem}
\usepackage{wasysym}
\usepackage{amssymb}
\usepackage{graphicx}
\usepackage{float}
\usepackage{adjustbox}
\usepackage{comment}
\usepackage{lscape}

\usepackage[utf8]{inputenc}

\def\be{\begin{equation}}
\def\ee{\end{equation}}
\def\ba{\begin{eqnarray}}
\def\ea{\end{eqnarray}}
\def\m{\mathrm}

\def\Mdotstar{\dot{M}_\ast}

\def\Omegadot{\dot{\Omega}}

\def\Mdotin{\dot{M}_{\mathrm{in}}}

\def\Mdot{\dot{M}}
\def\Edot{\dot{E}}
\def\Pdot{\dot{P}}

\def\Msun{M_{\odot}}

\def\Lcool{L_{\mathrm{cool}}}
\def\Lacc{L_\mathrm{acc}}
\def\Mdotstar{\dot{M}_\ast}

\def\rlc{r_{\mathrm{LC}}}
\def\rout{r_{\mathrm{out}}}
\def\rco{r_{\mathrm{co}}}

\def\Lx{L_{\mathrm{x}}}

\def\Md{M_{\mathrm{d}}}

\def\r_m{r_\mathrm{m}}
\def\v_esc{v_\mathrm{esc}}

\def\rA{r_{\mathrm{A}}}

\def\Tp{T_{\mathrm{p}}}

\def\Teff{T_{\mathrm{eff}}}

\def\Firr{F_{\mathrm{irr}}}

\def\P0min{P_{0,{\mathrm{min}}}}

\def\m{\mathrm}

\def\cs{c_{\mathrm{s}}}

\def\Alfven{Alfv$\acute{\mathrm{e}}$n~}
\def\418{SGR 0418+5729}
\def\142{AXP 0142+61}

\def\ergpers{erg~s$^{-1}$}

\def\spers{s~s$^{-1}$}

\def\B0-P{$B_0$~--~$P$}

\title[Is PSR J0726--2612 an XDIN Progenitor?]{Is PSR J0726--2612 a dim isolated neutron star progenitor? }

\author[\c{S}. \"{O}zcan et al.]{
\c{S}. \"{O}zcan,$^{1}$\thanks{E-mail: seydad@sabanciuniv.edu}
A. A. Gen\c{c}ali,$^{1}$
\"{U}. Ertan,$^{1}$
\\
$^{1}$Sabanc{\i} University, Orhanl{\i} Tuzla, 34956 \.{I}stanbul, Turkey
}

\date{Accepted XXX. Received YYY; in original form ZZZ}

\pubyear{2020}

\begin{document}
\label{firstpage}
\pagerange{\pageref{firstpage}--\pageref{lastpage}}
\maketitle

\begin{abstract}
The rotational properties and X-ray luminosity of PSR J0726--2612 are close to those of dim isolated neutron stars (XDINs). 
It was proposed that the source could be the first XDIN with observable pulsed radio emission. We have investigated the long-term evolution of the source to test this possibility in the fallback disc model.
Reasonable model curves that can account for the evolution of PSR J0726--2612 consistently with its radio pulsar property are similar to those of high-B radio pulsars with dipole field strength $B_0 \sim$ a few $\times 10^{12}$~G at the pole of the star.
In the same model, XDINs are estimated to have relatively weak fields ($B_0 \lesssim 10^{12}$~G) locating them  well below the pulsar death line.
From the simulations, we estimate that PSR J0726--2612 is at an age of $ t \sim 5 \times 10^4 $~yr, and  will achieve the rotational properties of a normal radio pulsar within $\sim 10^5$~yr, rather than the XDIN properties. 
\end{abstract}




\section{Introduction}

    X-ray dim isolated neutron stars (XDINs) form an isolated neutron star population among other young neutron star systems, namely anomalous X-ray pulsars (AXPs), soft gamma ray repeaters (SGRs), rotating radio transients (RRATs), high-B radio pulsars (HBRPs), and central compact objects (CCOs). At present, there are seven known XDINs characterized by their thermal X-ray emission with blackbody temperatures ranging from 40 to 110 eV and low X-ray luminosities in the range of $10^{31}-10^{32}$~erg~s$^{-1}$ \citep{Haberl2007, Turolla2009, Kaplan2011}.  Rotational periods of XDINs are in the $3$~--~$17$~s range \citep{Haberletall2004, HaberlF2004, Tiengo2007, Kaplan2009a, Kaplan2009b, Hambaryan2017} similar to those of AXPs and SGRs. Their characteristic ages $\tau_c=P/{2\dot{P}}=(1-4)\times 10^6$~yr where $P$ and $\Pdot$ are the rotational period and the period derivative of the neutron star. The kinematic ages are estimated to be between a few $10^5$~yr and $10^6$~yr by \citet{Sepeagle2011} which are consistent with the estimated cooling ages of the sources \citep{Page2009}. Soft gamma bursts, shown by AXPs and SGRs, continuous pulsed radio emission, or short radio bursts seen from RRATs have not been observed from XDINs \citep{Mereghetti2011a}. 
	
	With the assumption that XDINs evolve with purely magnetic dipole torques, the dipole field strengths are inferred to be $B_0=6.4\times10^{19}(P\dot{P})^{1/2} \sim $ a few $10^{13}$~G at the poles of the sources. 
	These strong dipole fields place XDINs above the pulsar death line in the \B0-P plane \citep{Haberl2007}, while no pulsed radio emission has been detected from these sources yet. It was proposed that the non-detection of pulsed radio emission from XDINs could be due to narrow beaming of their radio emission \citep{Haberl2005}. Recently observed radio pulsar PSR J0726--2612 (hereafter J0726) was proposed to be a good candidate to be an XDIN with an observable radio beam \citep{Rigoselli2019}. For this source, $P=3.44$~s, close to the minimum of XDIN periods, and $\Pdot= 2.93\times 10^{-13} $~s~s$^{-1}$, which give a characteristic age of $\sim 2\times 10^5$~yr. 
	The distance estimated from the dispersion measure \citep{Burgay2006} using the electron distribution model of \citet{Yao2017} gives $d \sim 3$~kpc. Nevertheless, the dispersion measure is likely to be effected by the Gould Belt \citep{Popov2005} crossing the line of sight to J0726. If the source is located in the Gould Belt as suggested by \citet{Sepeagle2011}, then $d \lesssim 1$~kpc. For the model calculations, we take $d = 1$~kpc which gives an X-ray luminosity $\Lx \simeq 4 \times 10^{32}$~erg~s$^{-1}$ and comparable to the rotational power $\Edot=I\Omega_\ast \Omegadot_\ast=2.8\times 10^{32}$~erg~s$^{-1}$ of the source \citep{Rigoselli2019,Vigano2013}, where $I$ is moment of inertia, $\Omega_\ast$ is the angular velocity of the neutron star and $\Omegadot_\ast$ is its time derivative. 
	Since the rotational properties and $\Lx$ of J0726 are in the ranges of those observed from HBRPs ($P = 0.15$~--~$6.7$~s, $\Pdot = 2.33 \times 10^{-14}$~--~$4.02 \times 10^{-12}$~\spers, $\Lx \simeq 10^{31}$~--~$4 \times 10^{34}$~\ergpers), the source is also classified as a HBRP \citep{Sepeagle2011,Olausen2013,Watanabe2019}.
	
	After a supernova explosion, a fallback disc can be formed around the neutron star \citep{Colgate1971,Michel1988, Chevalier1989, Perna2014}. 
	To explain the properties of AXPs,  \citet{Chatterjee2000} proposed that these sources are evolving with fallback discs. It was proposed that emergence of different isolated neutron star populations could be explained if the properties of fallback discs are included in the initial conditions together with initial period and magnetic dipole moment \citep{Alpar2001}. 
	Fallback discs were invoked to explain different rotational characteristics of isolated neutron stars that are not explained by evolutions with purely dipole torques \citep{Marsden2001, Menou2001, Eksi2003, Yan2012, Lei2013}. Emission properties of the fallback discs were also studied extensively \citep{Perna2000, Ertan2007}. It was shown by \citet{Ertan2007} that the broad-band spectrum of 4U 0142+61 from the optical to mid-IR bands \citep{Hulleman2000, Hulleman2004, Morii2005, Wang2006} can be accounted for by the emission from the entire disc surface. The fallback disc model proposed by \citet{Alpar2001} was developed later including the effects  of the X-ray irradiation, cooling luminosity, and inactivation of the disc in the long-term evolution \citep{ErtanE2009, Ertan2014}.
	When there is a fallback disc around the star, the spin-down torque arising from the interaction of the inner disc with the dipole field of the star usually dominates the magnetic dipole torque.  In the fallback disc model, $B_0$ values are estimated to be one to three orders smaller than the values inferred from the dipole torque formula. The long-term evolution of XDINs and HBRPs with fallback discs was studied earlier by \citet{Ertan2014} and \citet{Benli2017, Benli2018}.
	The model can reproduce $P$, $\Pdot$ and $\Lx$ of individual XDIN and HBRP sources with $B_0$ in the ranges of ($0.3$~--~$1.3$) $\times 10 ^{12}$~G for XDINs and ($0.3$~--~$6) \times 10^{12}$~G for HBRPs. These relatively weak fields together with the long periods place XDINs well below the pulsar death line \citep[see][figure 4]{Ertan2014} in the $B_0-P$ diagram \citep[][]{Chen1993}, while HBRPs with relatively strong fields and/or short periods are located above the death line \citep{Benli2017, Benli2018}.
	In other words, in the fallback disc model, the lack of radio pulses from XDINs is due to their weak dipole fields, rather than the beaming geometry.
	
	In this work, our aim is to investigate the long-term evolution of J0726, and compare its properties and evolution with those of XDINs and HBRPs in the fallback disc model. The same model was applied earlier to AXP/SGRs, CCOs, and RRATs as well \citep{Ertan2007,ErtanE2009,Ertan2014,Caliskan2013,Benli2017,Gencali2018}. In Section 2, we briefly describe this model. We discuss the  results of model calculations in Section 3, and summarize our conclusions in Section 4.

\section{The Model}

\label{secconc}
	
	Since the details of the model calculations and its applications to different neutron star populations are given in earlier works \citep[see e.g.][]{ErtanE2009, Ertan2014, Benli2016},  we briefly describe the model calculations here.
	
	We solve the disc diffusion equation starting with a surface density profile of a steady disc using the kinematic viscosity $\nu = \alpha \cs h$ where, $\alpha$ is the kinematic viscosity parameter, $\cs$ is the local sound speed, and $h$ is the pressure scale height of the disc \citep{Shakura1973}. 
	In the accretion with spin-down (ASD) phase, we calculate the disc torque, $N_\m{D}$, acting on the star by integrating the magnetic torques from the conventional \Alfven radius, $\rA \simeq (GM)^{-1/7}\mu^{4/7}\Mdotin^{-2/7} $ \citep{Lamb1973, Davidson1973}, to the co-rotation radius, $\rco = (GM/\Omega_\ast^2)^{1/3}$ where $G$ is the gravitational constant, $M$ is the mass of the neutron star and $\mu$ is its magnetic dipole moment, $\Mdotin$ is the mass inflow-rate  at the inner disc. 
	The magnitude of this torque could be written in terms of $\Mdotin$ and $\rA$ as $N_\m{D} = \frac{1}{2} \Mdotin (GM \rA)^{1/2} \Big[ 1 - (\rA/\rco)^3 \Big]$ \citep[see][for details]{Ertan2008}.
	The contributions of the magnetic dipole torque, $N_\m{dip}$, and the spin-up torque associated with accretion on to the star, $N_\m{acc}$, are negligible in the long-term accretion regime of XDINs \citep{Ertan2014}. That is, the total torque acting on the star $N_\m{TOT} = N_\m{D} + N_\m{dip} + N_\m{acc}$ is dominated by $N_\m{D}$ in the ASD phase of XDINs. In this regime, $\rco < \rA < \rlc$, where $\rlc=c/\Omega_\ast$ is the light cylinder radius, and $c$ is the speed of light. 
	
	During the ASD phase, $\rA$ increases with gradually decreasing $\Mdotin$, and eventually  becomes equal to $\rlc$. For $\Mdotin$ below this critical value, we replace $\rA$ with $\rlc$ in the $N_\m{D}$ equation. In the model, $\rA = \rlc$ is also the condition for the propeller-accretion transition. Since $\Mdotin$ enters a sharp decay phase at the end of the ASD phase, exact value of $\Mdotin$ for the accretion-propeller transition does not affect our results significantly. In the strong-propeller (SP) phase, we assume that all the matter inflowing to the inner disc is expelled from the system. The pulsed radio emission is allowed only in the SP phase when there is no accretion on to the source. 
	
	In the ASD phase, the mass accretion on to the star produces an X-ray luminosity, $\Lacc = G M \Mdot_\ast/R_\ast $, where $R_\ast$ is the radius of the neutron star. In this phase, we take the mass accretion rate $\Mdotstar = \Mdotin$. The total X-ray luminosity, $\Lx = \Lacc + \Lcool$, where $\Lcool$ is the intrinsic cooling luminosity of the star \citep{Page2009}.  
	In the $\Lcool$ calculation, we also include the small contribution of the external torques to the internal heating of the neutron star \citep{Alpar2007}.
	In the SP phase, $\Mdot_\ast = 0$, $N_\m{acc} = 0$, and $\Lx = \Lcool$, since accretion is not allowed in this regime. The disc is heated by X-ray irradiation in addition to the viscous dissipation. The effective temperature of the disc can be written as $\Teff \simeq \Big[(D + \Firr) /\sigma \Big]^{1/4}$ where $D$ is the rate of viscous dissipation per unit area of the disc, $\sigma$ is the Stefan-Boltzmann constant, $\Firr = 1.2 C \Lx/ (\pi r^2)$ is the irritation flux, where $r$ is the radial distance from the star, $C$ is the irradiation parameter, which depends on the albedo and geometry of the disc surfaces \citep{Fukue1992}. The disc becomes viscously inactive below a critical temperature, $\Tp$. Dynamical outer radius, $\rout$, of the viscously active disc is equal to the radius currently at which $\Teff = \Tp$. Across the outer disc, $\Firr$ dominates D, that is, the X-ray irradiation significantly affects the long-term evolution of the source by extending the life-time of the active disc.
	
	The main disc parameters ($\alpha$, $C$, $\Tp$) for the fallback discs of different neutron star populations are expected to be similar. The same model employed here can reproduce the individual source properties of AXP/SGRs, CCOs, HBRPs and XDINs with $\Tp \sim$~($50$~--~$150$)~K, and C = ($1$~--~$7) \times 10^{-4}$ \citep{Ertan2006, Ertan2007, Caliskan2013, Ertan2014, Benli2016, Benli2017, Benli2018, BenliCCO2018}. These $\Tp$ values in the model are in good agreement with the results of the theoretical work indicating that the disc is likely to be active at very low temperatures \citep{Inutsuka2005}. The range of $C$ estimated in our model is similar to that estimated from the optical and X-ray observations of the low-mass X-ray binaries \citep[see e.g.][]{Dubus1999}. We try to obtain the properties of J0726 also with these main disc parameters. 
	This provides a systematic comparison between the initial conditions of different populations, namely the magnetic dipole field strength $B_0$, the initial disc mass $\Md$, and the initial period $P_0$.
	
	The $\alpha$ parameter does not significantly affect the long-term evolution. The conditions in a slowly evolving fallback disc are similar to steady-state conditions.  The outer regions of the active disc govern the rate of mass-flow to the inner disc. That is, the $\alpha$ parameter in our model should be considered as the property of the outer disc. $\Tp$ and $C$ are degenerate parameters. With smaller $\Tp$ values the active disc has a longer lifetime. A stronger irradiation (greater $C$) also extends the lifetime of the active disc. A detailed discussion about the effects of these parameters on the evolution of the neutron star can be found in \citet{ErtanE2009}.

\section{Results and Discussion}
	
\begin{figure}
		\centering
		\includegraphics[width=\columnwidth]{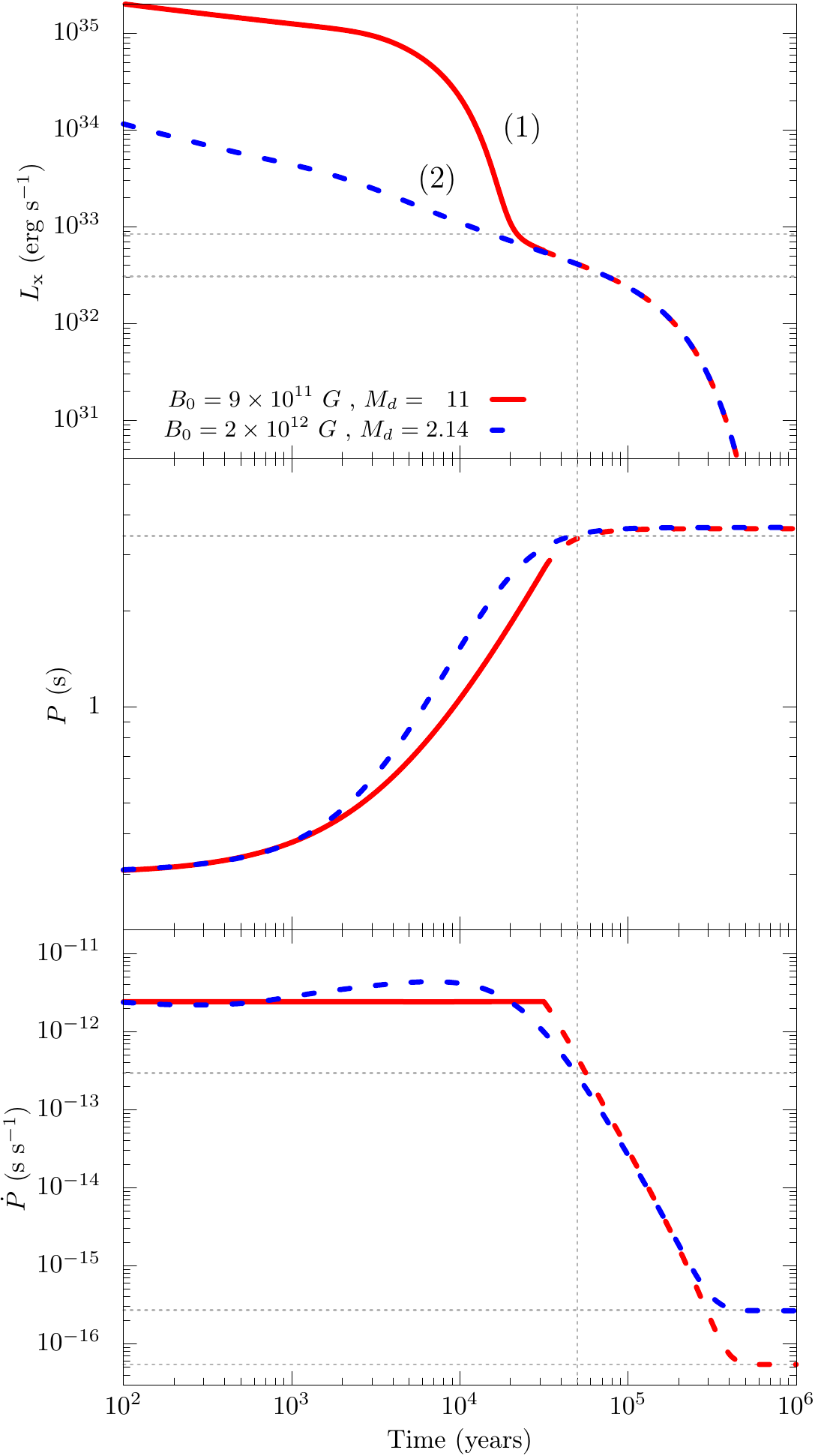}	
		\caption{
			Illustrative model curves for the long-term evolution of PSR J0726--2612. 
			The curves are obtained with $B_0$ and $\Md$ (in units of $10^{-6}~\Msun$) values given in the top panel. 
			The main disc parameters employed in both models are $C=1 \times 10^{-4}$, $\Tp=100$~K, and $\alpha = 0.045$.
			Horizontal lines show the observed $P=3.44$~s, $\Pdot=2.93 \times 10^{-13}$~s~s$^{-1}$, and the estimated $\Lx$ range for $d=1$~kpc \citep{Rigoselli2019}.
			For curve 1, solid and dashed branches correspond to the ASD and SP phases respectively. For the evolution represented by curve 2, the source always remains in the SP phase, and this curve is a more likely representation of the evolution of PSR J0726--2612 (see the text for details). Eventually, $\Pdot$ curves converge to the levels corresponding to the magnetic dipole torques (shown by two horizontal dotted lines at the bottom of the $\Pdot$ panel).
		}
		\label{fig:plot}
	\end{figure}

	The model curves seen in Fig. \ref{fig:plot} illustrate two different evolutionary histories for J0726: 
	(1) The model source following curve 1 starts its evolution in the ASD phase with $\Lx \simeq \Lacc$, and remains in this phase until it makes a transition to the SP phase at $t \sim 3 \times 10^4$~yr. 
	The solid and dashed branches of the curves correspond to the ASD and SP phases respectively.
	It is seen in the middle and bottom panels that the rapid increase of $P$ stops with sharply decreasing $\Pdot$ after the ASD/SP transition.
	(2) Curve 2 represents the evolution of a neutron star that remains always in the SP phase with $\Lx \simeq \Lcool$. For a given $B_0$, the sources with $\Md$ smaller than a critical value cannot enter the ASD phase, and evolves in the SP phase likely as radio pulsar.
	The rotational evolution for this type of evolution is sensitive to $\Md$, while for type (1) solution, $P$ and $\Pdot$ evolution do not significantly depend on $\Md$ \citep[see e.g.][]{Benli2016}.
	
	\begin{figure}
		\centering
		\includegraphics[width=\columnwidth]{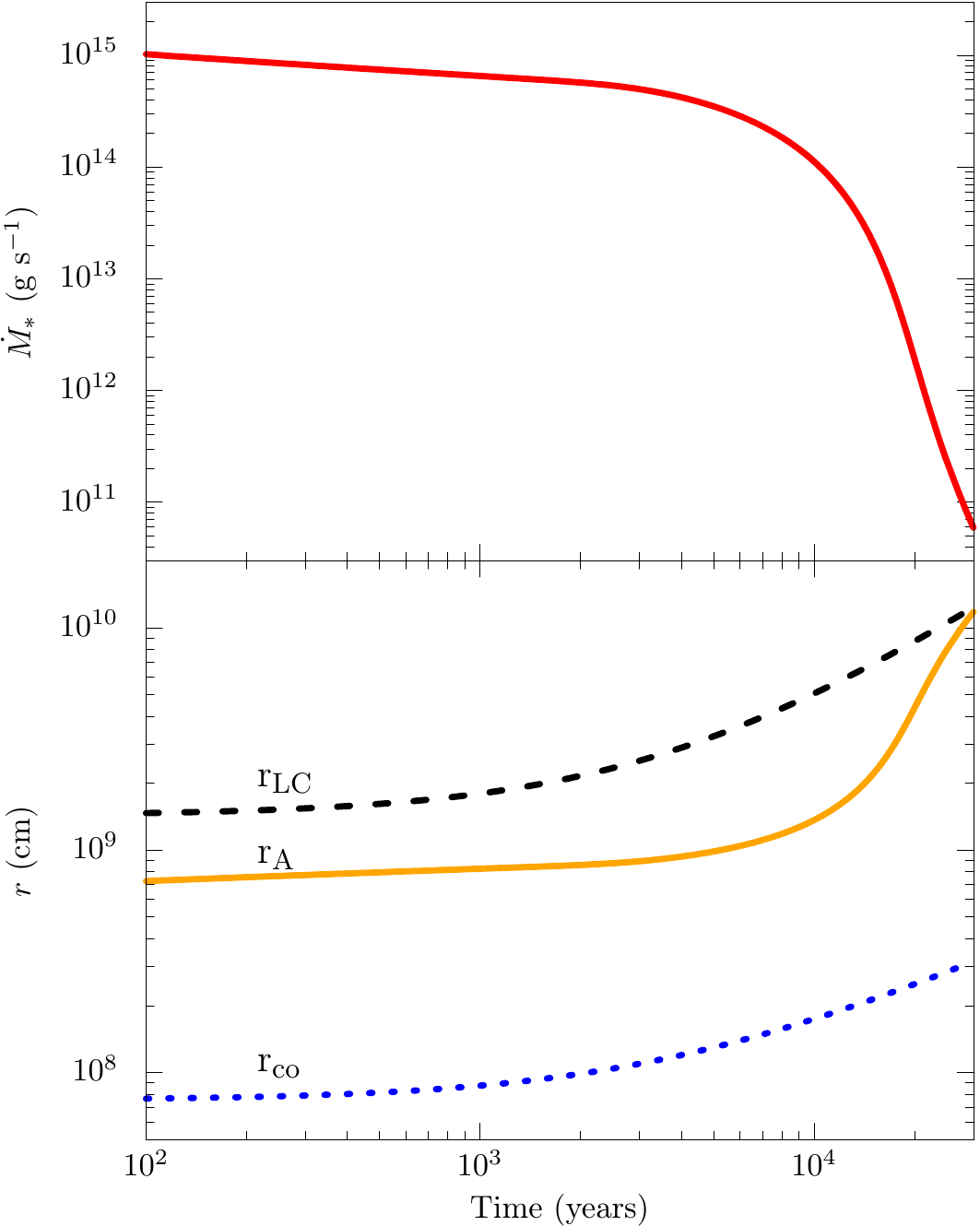}	
		\caption{The evolution of the accretion rate, $\rco$, $\rA$ and $\rlc$ in the ASD phase of type (1) evolution (see Fig. \ref{fig:plot}). The accretion is switched off at $t \simeq  3 \times 10^4$~s, and the system enters the SP phase (see the text).}
		\label{fig:Mdot_ra}
	\end{figure}
	
\begin{figure}
	\centering
	\includegraphics[width=\columnwidth]{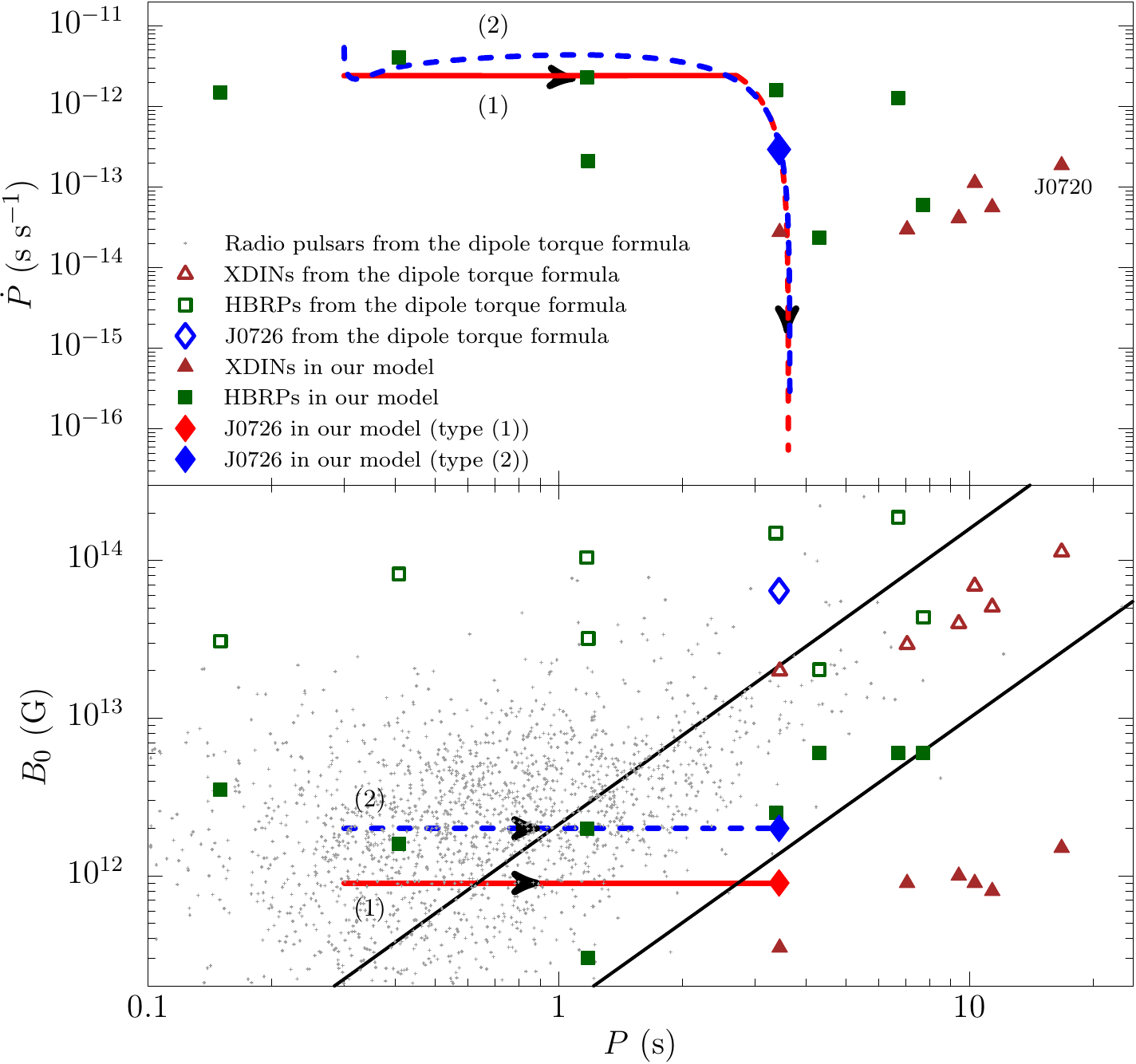}
	\caption{
		Long-term evolution in the $P~-~\Pdot$ and $B_0-P$ diagrams for the same model curves given in Fig. \ref{fig:plot}. XDINs and HBRPs are indicated by triangles and squares respectively. In the $B_0-P$ plane, empty  symbols show $B_0$ values inferred from the dipole torque formula using $P$ and $\Pdot$ values \citep[ATNF Pulsar Catalogue version 1.63,][]{Manchester2005}\protect\footnotemark. The filled symbols indicate the average $B_0$ values estimated in our model \citep{Ertan2014,Benli2017,Benli2018}. The solid lines are the upper and lower borders of the pulsar death valley \citep{Chen1993}. The filled diamonds show the current location of J0726 estimated for type (1) and type (2) solutions.	}
	\label{fig:B0-P}
	\end{figure}

	Illustrative model curves seen in Fig. \ref{fig:plot} are obtained with the main disc parameters: $\alpha=0.045$, $\Tp = 100$~K, $C =1 \times 10^{-4}$. The initial conditions for curve 1 are
	$P_0=0.3$~s, $\Md = 1.1 \times 10^{-5}~\Msun$, $B_0 = 9 \times 10^{11}$~G. The maximum $B_0$ allowed for the type (1) solution (curve 1) is $\sim 1.2 \times 10^{12}$~G, while the type (2) solution can reproduce the source properties with $B_0 \gtrsim 1.5 \times 10^{12}$~G. For $P = 3.44$~s, the minimum $B_0$ required for the pulsed radio emission is $\sim 1.4 \times 10^{12}$~G. 
	In Fig. \ref{fig:Mdot_ra}, we have also plotted the evolution of $\Mdotstar$, $\rA$, $\rco$ and $\rlc$ in the ASD phase of type (1) solution.
	Due to the simplifications in our model, we cannot exclude type (1) evolution.
	Nevertheless, even if the source is inside the death valley, it is too close to the death line, which implies that this solution is not very likely to represent the actual evolution of J0726. 
	Most of the radio pulsars seem to die inside the death valley at points not very close to the pulsar death line. Otherwise, if the sources switch off the radio pulses when crossing the death line, their number density would increase close to the death line, which is not observed. Some of the sources die close to the upper boundary, while some others close to the lower boundary (death line), altogether forming a roughly homogeneous distribution inside the death valley. 
	For our type (1) solution,  after termination of the ASD phase, the source finds itself very close to the lower boundary. In this case, the star can show radio pulses only if its actual death point is indeed very close to the lower boundary.
	Type (2) solution seems to be more reasonable representation of the evolution of J0726. This evolution is similar to those of some of the HBRPs in the same model \citep{Benli2017, Benli2018}.
	For both solutions, the source is currently evolving in the SP phase at an age $\sim 5 \times 10^4$~yr.
	At present, the star is slowing down dominantly by the disc torque that will eventually decrease below the magnetic dipole torque at $t \sim$~a few ~$\times~10^5$~yr. For instance, for $B_0 = 2 \times 10^{12}$~G, (curve 2) the sharp decrease in $\Pdot$ will continue down to $\Pdot \simeq 2.7 \times 10^{-16}$~s~s$^{-1}$.
	Our results indicate that J0726 will evolve to these ages with a $\Pdot$ that is about three orders of magnitude smaller than its present value (Fig. ~\ref{fig:plot}). This means that the source is likely to be classified as a normal radio pulsar  with $B_0 \sim $ a few $\times 10^{12}$~G deduced from $P$ and $\Pdot$ at the ages of XDINs.
	In Fig. \ref{fig:B0-P}, we have plotted the evolution of J0726 in the $P~-~\Pdot$ and $P~-~B_0$ diagrams together with XDINs and HBRPs with the properties estimated in our model.
	
    \footnotetext{\url{https://www.atnf.csiro.au/research/pulsar/psrcat/}}
	
		\begin{figure}
		\centering
		\includegraphics[width=\columnwidth]{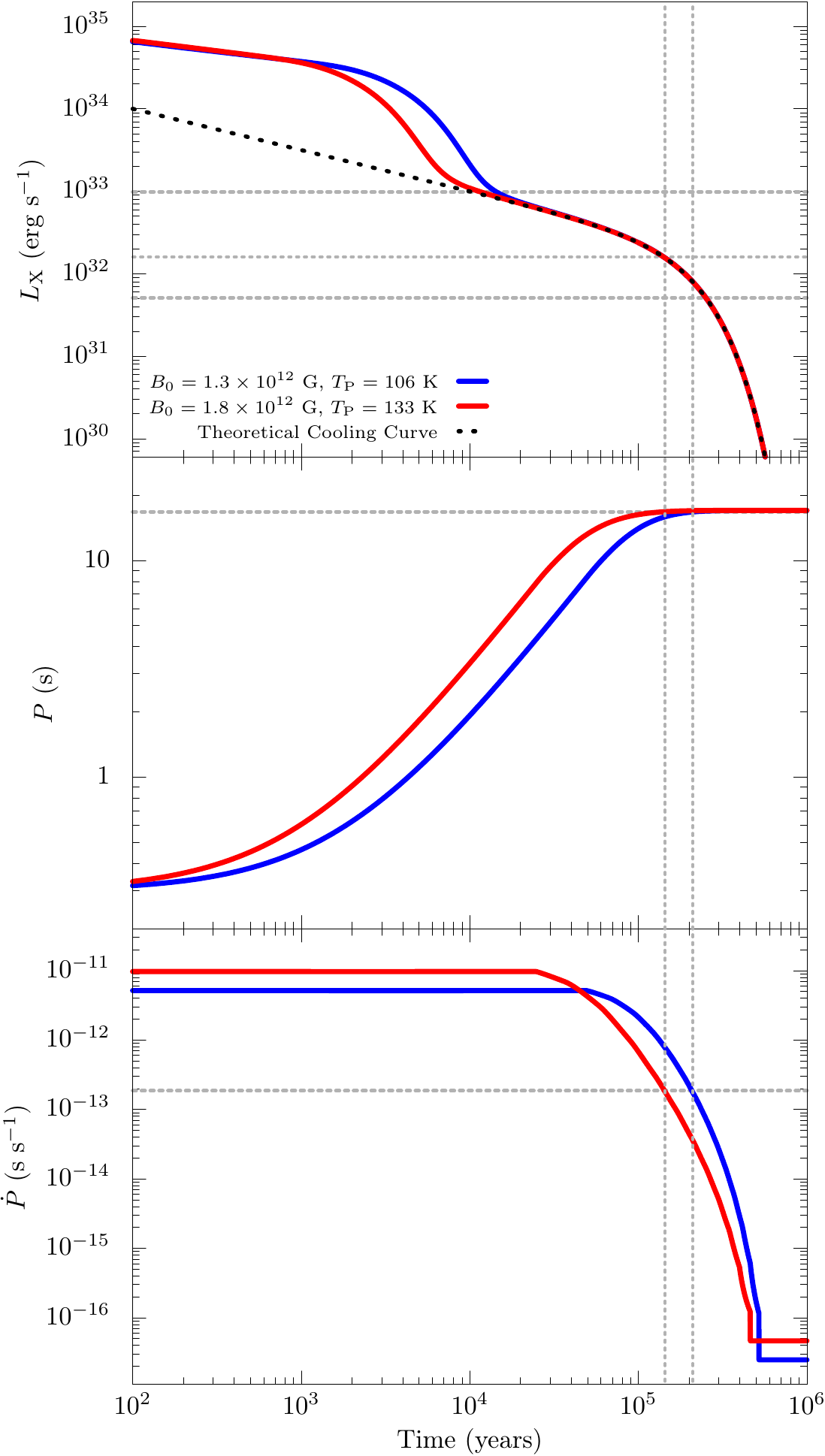}	
		\caption{Illustrative model curves for the long-term evolution of RX J0720.4–-3125 with the updated period and period derivative. For both models, $\alpha = 0.045$, $C = 1 \times 10^{-4}$, $P_0 = 0.3$~s, $\Md = 4.74 \times 10^{-6} \Msun$. The curves obtained with $B_0$ and $\Tp$ values given in the top panel. The dotted curve indicates the theoretical cooling curve \citep{Page2009}. Horizontal dashed lines show $P = 16.78$~s, $\Pdot = 1.86 \times 10^{-13}$~s~s$^{-1}$, $\Lx = 1.6 \times 10^{32}$ erg~s$^{-1}$ as used in \citet{Ertan2014} assuming $d=270$~pc.  There is a large uncertainty in $d = 280^{+210}_{-85}$~pc \citep{Eisenbeiss2011, Tetzlaff2011, Hambaryan2017}.}
		\label{fig:0720}
	\end{figure}
	
	In our present and earlier works, we employed theoretical cooling curve estimated by \citet{Page2009} for conventional dipole fields. This cooling curve could differ from the actual cooling curve depending on some unknown details of the neutron star properties like equation of state and mass of the star \citep[see e.g.][]{Potekhin2018, Potekhin2020}. For the sources that are currently in the ASD phase, the details of the cooling curve do not affect our results, but could modify our model parameters for sources in the SP phase, like XDINs. The ages of XDINs estimated in our model are on the average a few times smaller than the estimated kinematic ages. If the actual ages are indeed close to the kinematic ages, the source properties can be obtained with $B_0$ values smaller than we reported here and in \citet{Ertan2014} by a factor smaller than two. The field strengths  estimated in our model should be considered taking these uncertainties into account.  These small changes in $B_0$ do not change the qualitative features of the model curves for XDINs.
    Recently, the period of RX J0720.4--3125 was updated from $8.39$~s to $16.78$~s by \citet{Hambaryan2017}. The period derivative of the source was also updated from $\sim 7 \times 10^{-14}$~s~s$^{-1}$ to  $1.86 \times 10^{-13}$~s~s$^{-1}$ by \citet{HambaryanNeuh2017}. 
    For this source, we have performed new simulations and modified the model parameters obtained by \citet{Ertan2014}. Our results  and model parameters are given in Fig. \ref{fig:0720}. With the updated period and period derivative, using the same main disc parameters as employed in \citet{Ertan2014}, the model can reproduce the source properties with slightly higher $B_0$ ($(1.3-1.8) \times 10^{12}$~G) values in comparison with the $B_0$ obtained in \citet{Ertan2014}. Similar results could be produced for a large range of disc masses.

    In our model, the inner disc interacts with the large-scale magnetic dipole field of the neutron star. Close to the surface of the star, there could be quadrupole magnetar fields which could affect the surface temperature distribution and the absorption features \citep{Guver2011}. 
    Presence of these small-scale strong fields in XDINs and other isolated neutron star populations is compatible with the fallback disc model, nevertheless these detailed surface properties are not addressed in our long-term evolution model. Some other spectral features that could be produced by the disc-field interaction should also be studied independently. In particular, \citet{Ertan2017} showed that the heating of the inner disc boundary by the magnetic stresses can account for the optical/UV excesses of XDINs, while the entire disc spectra are consistent with the observed upper limits in the IR bands. We note that there is an uncertainty in the disk spectrum because of unknown
    inclination angle of the disk. To estimate the entire disk spectrum, at least a single detection in one of the IR bands is needed. At present, there is no IR/optical detection or upper limits estimated for J0726.

    The X-ray luminosities of XDINs exceed their spin-down powers. In our long-term evolution model, this is a natural outcome of rapid increase in periods by efficient disc torques and sharp decrease in $\Pdot$ in the late SP phase, which leaves the observed spin-down powers below the cooling luminosities of these sources. The current periods of XDINs together with the weak fields estimated in our model place these sources below the pulsar death line \citep[][plotted also in Fig. \ref{fig:B0-P}]{Ertan2014}. This indicates that known XDINs cannot emit radio pulses in our model. There are a few exceptional active radio pulsars that are close to, but below the death line, namely PSR J0250+5854 with $P = 23.5$~s \citep{Tan2018}, PSR J2251--3711 with $P = 12.1$~s \citep{Morello2020}, PSR J2144--3933 with $P = 8.5$~s \citep{Young1999}. Nevertheless, in our model, we find the locations of XDINs well below the death line where there are no radio pulsars.

	Is it possible that a source with $B_0 \sim$ a few $\times 10^{12}$~G starts in the ASD phase  with a greater $\Md$?  It is possible, and we estimate that these sources become AXP/SGRs, and evolve to relatively long periods which leave them below the pulsar death line at the end of the ASD phase. In our model, these sources can never become radio pulsar in their lifetimes provided that the accretion is not hindered occasionally due to instabilities at the inner disc.

\section{Conclusions}
	
	We have shown that $P$, $\Pdot$ and $\Lx$ of J0726 can be achieved by a neutron star evolving with a fallback disc. We have found that there are two possible evolutionary histories that could produce the properties of J0726. For both solutions, the source is in the strong-propeller (SP) phase at present. 
	In the first alternative (curve 1 in Fig. ~\ref{fig:plot}), the star initially evolves in the accretion with spin-down (ASD) phase, and makes a transition into the SP phase at an earlier time of its evolution. 
	For the second type of solution (curve 2 in Fig. ~\ref{fig:plot}), the source always evolves in the SP phase. Since the X-ray luminosity is powered by the cooling luminosity in the SP phase, the model sources reach the properties of J0726 at an age close to the estimated cooling age ($\sim 5 \times 10^4$~yr) of J0726.
	
	The radio pulsars following the type (1) evolution are not likely to be common, since these sources find themselves very close to the pulsar death line after the accretion is switched off. The curve 2 seems to show a more likely evolution for J0726 which is also similar to the evolution of some of the HBRPs, rather than XDINs. The model curves indicate that the source will acquire the rotational properties of normal radio pulsars at the ages of XDINs (Fig. ~\ref{fig:plot}).

	In our long-term evolution model, the basic difference between the HBRPs and XDINs are the field strengths $B_0$. XDINs with relatively small $B_0$ ($10^{11}~-~10^{12}$~G) tend to start their evolution in the ASD phase, since it is easier for the inner disc to extend down to the co-rotation radius for weaker dipole fields. On the other hand, HBRPs with stronger fields ($B_0 \gtrsim 2 \times 10^{12}$~G) either always evolve in the SP phase, as we estimate for J0726, or make a transition from the initial SP phase to the ASD phase at a later time of evolution. In the latter case, the sources are expected to evolve to the properties of AXP/SGRs \citep{Benli2017}. A detailed comparison of the long-term evolutions and the statistical properties of these neutron star populations in the fallback disc model will be studied in an independent work.
	
\section*{Acknowledgements}

	We thank the referee for useful comments that have improved this manuscript. We acknowledge research support from Sabanc{\i} University, and from T\"{U}B\.{I}TAK (The Scientific and Technological Research Council of Turkey) through grant 117F144.




\bibliographystyle{mnras}
\bibliography{mnras.bib} 

\bsp	
\label{lastpage}
\end{document}